\documentclass[aps, twocolumn, superscriptaddress, floatfix]{revtex4-1}

\newcommand{\flip}{{\mathrm{flip}}}
\newcommand{\IR}{{\mathrm{\scriptscriptstyle IR}}}
\usepackage{hyperref}
\hypersetup{
	colorlinks=true,
	linkcolor=blue,
	filecolor=magenta,      
	urlcolor=blue,
	citecolor=blue,
}
\usepackage{float}
\usepackage{amsmath,bm,bbm}
\usepackage{amssymb}
\usepackage{cleveref}

\usepackage{enumitem}
\usepackage{textcomp}
\usepackage{feynmp}
\usepackage{graphicx}

\usepackage{dsfont}
\usepackage{color}
\usepackage[makeroom]{cancel}

\definecolor{purple}{rgb}{0.8,0,0.6}

\renewcommand\thesection{\arabic{section}}
\renewcommand\thesubsection{\thesection.\arabic{subsection}}
\renewcommand\thesubsubsection{\thesubsection.\arabic{subsubsection}}
\renewcommand\theparagraph{\thesubsubsection.\arabic{paragraph}}
\renewcommand\thesubparagraph{\theparagraph.\arabic{subparagraph}}

\renewcommand\appendix{\par
  \setcounter{section}{0}%
  \setcounter{subsection}{0}%
  \setcounter{figure}{0}%
  \renewcommand\thesection{\Alph{section}}
  \renewcommand\thesubsection{\thesection.\arabic{subsection}}
  \renewcommand\thesubsubsection{\thesubsection.\arabic{subsubsection}}
  \renewcommand\theparagraph{\arabic{paragraph}}
  \renewcommand\thesubparagraph{\arabic{subparagraph}}
}

\begin{document}
\title{Evolution of the Primordial Axial Charge across Cosmic Times}

\author{A.~Boyarsky}
\affiliation{Instituut-Lorentz for Theoretical Physics, Universiteit Leiden, Niels Bohrweg 2, 2333 CA Leiden, Netherlands}

\author{V.~Cheianov}
\affiliation{Instituut-Lorentz for Theoretical Physics, Universiteit Leiden, Niels Bohrweg 2, 2333 CA Leiden, Netherlands}

\author{O.~Ruchayskiy}
\affiliation{Niels Bohr Institute, University of Copenhagen, Blegdamsvej 17, DK-2100 Copenhagen, Denmark}

\author{O.~Sobol}
\affiliation{Institute of Physics, Laboratory for Particle Physics and Cosmology, \'{E}cole Polytechnique F\'{e}d\'{e}rale de Lausanne, CH-1015 Lausanne, Switzerland}
\affiliation{Physics Faculty, Taras Shevchenko National University of Kyiv, 64/13, Volodymyrska Str., 01601 Kyiv, Ukraine}
\email{oleksandr.sobol@epfl.ch}

\begin{abstract}
We investigate collisional decay of the axial charge in an
electron-photon plasma at temperatures 10~MeV--100~GeV.
We demonstrate that the decay rate of
the axial charge is first order in the fine-structure constant $\Gamma_{\rm flip}\propto \alpha m_{e}^{2}/T$ and thus
orders of magnitude greater than the naive 
estimate which has been in use for decades. 
This counterintuitive result arises through infrared divergences
regularized at high temperature by environmental effects.
The decay of axial charge plays an important role in the problems of leptogenesis and cosmic magnetogenesis. 
\end{abstract}

\maketitle

The origin of cosmic magnetic fields remains a subject of intense debate, see Refs.~\cite{Grasso:2000wj,Subramanian:2009fu,Kandus:2010nw,Durrer:2013pga} for reviews. 
A leading hypothesis is that these fields originated in the hot and homogeneous early Universe. 
If this hypothesis is correct, 
the requirements that the magnetic fields (i) were germinated before
and (ii) survived until the beginning of the structure formation epoch (when the process of their amplification started)---impose tight constraints on the possible history of the Universe, likely implying the existence of new physics \cite{Durrer:2013pga,Hortua:2019apr}. This potential for serving as a bridge between the observational data and the properties of the early Universe makes both primordial magnetogenesis and 
magnetohydrodynamics (MHD) of ultrarelativistic plasmas research 
topics of fundamental importance.
It has been argued that due to the weakness of nonconservation of the 
axial charge current in an ultrarelativistic plasma, the proper description of the evolution of primordial cosmic magnetic fields requires an extension of
MHD called \textit{chiral magnetohydrodynamics }~\cite{Joyce:1997uy,Boyarsky:2011uy,Rogachevskii:2017uyc}, see also Refs.~\cite{Giovannini:2013oga,DelZanna:2018dyb}.
In chiral MHD the system of Maxwell and Navier-Stokes equations is supplemented with an extra degree of freedom---the axial chemical potential. Such an 
extension materially affects the predictions of the theory. In particular, chiral MHD admits for the transfer of magnetic energy from short- to
long-wavelength modes of helical magnetic fields, partially compensating Ohmic
dissipation in the early Universe and thus increasing their chance to survive until today~\cite{Joyce:1997uy,Boyarsky:2011uy,Tashiro:2012mf,Hirono:2015rla,Dvornikov:2016jth,Gorbar:2016klv,Brandenburg:2017rcb,Schober:2018wlo}.
It is worth noting that chiral MHD has drawn a lot 
of recent interest not only because of its importance for the description of primordial magnetic fields, but also due to its relevance to the theory of neutron stars and quark-gluon plasmas (see, e.g., Refs.~\cite{Kharzeev:2011vv,Tashiro:2012mf,Boyarsky:2012ex,Akamatsu:2013pjd,Wagstaff:2014fla,Hirono:2015rla,Yamamoto:2015ria,Gorbar:2016klv,Long:2016uez,Pavlovic:2016gac,Dvornikov:2016jth,Gorbar:2016qfh,Sen:2016jzl,Brandenburg:2017rcb,Schober:2017cdw,Hirono:2017wqx,Hattori:2017usa,Gorbar:2017toh,Rogachevskii:2017uyc,Schober:2018ojn,Dvornikov:2018tsi,DelZanna:2018dyb,Schober:2018wlo,Masada:2018swb,Mace:2019cqo,Schober:2020ogz}).

The chiral MHD description is only appropriate inasmuch as the axial 
current can be treated as conserved on microscopic timescales such 
as the momentum and energy relaxation rates. 
This requires the typical kinetic energy of an electron in the plasma to 
significantly exceed the electron mass $m_e,$ so one can 
meaningfully assign chirality to each particle. In such a high-temperature 
regime, $T \gg m_e,$ the axial charge decays through rare chirality-flipping processes, which are still possible due to 
the nonconservation of chirality introduced by a perturbatively small
mass term. Surprisingly, the chirality flipping rate resulting from 
such processes has never been rigorously 
calculated \cite{Note1}.
The previous body of work relied on the naive 
estimate of the chirality flip rate 
\begin{equation}
  \label{eq:flip_naive}
  \Gamma_\flip^{\rm naive}
\propto \left(\frac{m_{e}}{T}\right)^{2}\alpha^{2}T
\end{equation}
as being second order in the 
small parameter responsible for chirality nonconservation  $m_e/T$  
and first order in the electron scattering rate $ \Gamma_{\rm scat}\propto \alpha^2 T$
(see, e.g., Refs.~\cite{Boyarsky:2011uy,Grabowska:2014efa,Manuel:2015zpa,Pavlovic:2016mxq}),
where $\alpha = e^2/(4\pi)$ is the fine structure constant.
This estimate is based on the simple rationale that for an ultrarelativistic particle in a definite helicity state, which up to a correction on the order of $m_e/T$ is the same as a definite chirality state, the helicity can only be flipped via a sideways scattering process having the rate $\Gamma_{\rm scat}$.

The aim of the present work is to show that contrary to the naive 
expectation, Eq.~\eqref{eq:flip_naive}, the actual chirality flipping rate in an ultrarelativistic plasma is first order in $\alpha$; see Eq.~\eqref{gamma-flip-kinetic-final}.
We focus, in particular, on the analysis of infrared singularities in the matrix elements of chirality-flipping Compton scattering and show how they effectively 
lead to the cancellation of one power of $\alpha.$ We also briefly discuss other scattering channels which contribute to chirality flipping in 
the same order of perturbation theory and give the 
resulting leading-order asymptotic expression for the chirality 
flipping rate. 
A detailed derivation of this results in the framework of
quantum field theory based linear response formalism can be found in our companion paper \cite{PaperII}. We note that although it is natural for kinetic coefficients associated with electron-photon scattering to be second order in the fine structure constant, there exists another known exception from this rule---the axial charge diffusion coefficient \cite{Hou:2017szz}.

\begin{figure}[ht!]
	\centering
	\includegraphics[height=3.5cm]{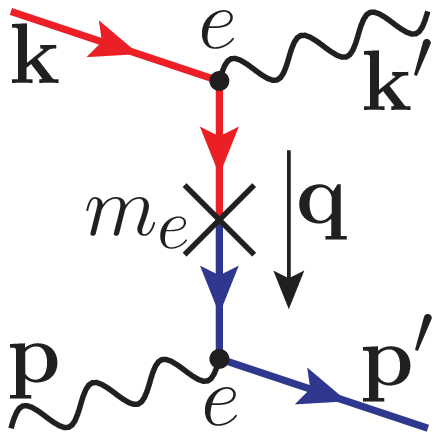} \hspace*{0.4cm}
	\includegraphics[height=3.5cm]{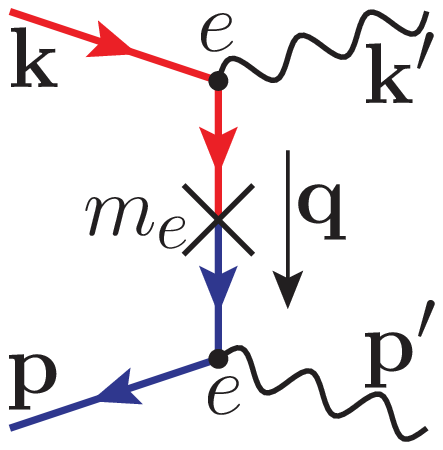}\\
	(a) \hfill (b)
	\caption{The $t$-channel Compton scattering (a) and electron-positron annihilation (b) with the chirality flip in the intermediate state contributing to the chirality equilibration rate. Although naively they are of the second order in $\alpha$, their amplitudes contain infrared singularities. Regularization of these singularities leads to the result which is of the first order in~$\alpha$. \label{fig-Compton}}
\end{figure}

Our main idea can be summarized as follows.
We consider $2\leftrightarrow 2$ chirality flipping processes, starting from the
massless QED limit and treating both the electron mass and the electron-photon 
coupling as perturbations
(see Fig.~\ref{fig-Compton}).
As is well known, such processes have a nonintegrable infrared singularity at small momentum transfer~\cite{Lee:1964is,Dolgov:1971ri}.
This signals the need for the resummation of the leading infrared divergence 
in all orders of the perturbation theory series. Such a resummation 
should generally result in an answer
$\Gamma_\flip \propto \alpha^2 T {m_e^2}/{q_\IR^2}$, where $q_\IR$ is the infrared regulator scale associated with either the effective mass or the 
lifetime of the quasiparticle associated with the electron propagator. 
In a hot plasma a natural infrared scale arises from the thermal self-energy
corrections to the dispersion relations of (quasi)particles. In the Supplemental Material~\cite{Suppl}, Sec.~A
we use hard thermal loops (HTL) resummation to show that 
such corrections are of 
the order $q_{\IR}\sim \sqrt\alpha T,$ which results in
$\Gamma_\flip \propto \alpha {m_e^2}/T.$
We note that such an approach is not valid in the regime 
where the self-energy corrections are less than 
the electron mass. Therefore the validity range
of our analysis is $T\ge m_e/\sqrt \alpha \sim 10$~MeV.

Next we describe our calculations in some detail.
Particle chiralities are well defined for free massless particles. 
Therefore we start from massless QED and treat mass as a perturbation.
In plasma this means that we consider each chirality obeying its own Fermi-Dirac distribution
\begin{equation}
\label{fermi-dirac}
f_{L,R}(\mathbf{k})=
\frac{1}{\exp[(\epsilon_{k}\pm\mu_5)/T]+1}\equiv n_{F}(\epsilon_{k}\pm\mu_5),
\end{equation}
with chemical potentials $\pm\mu_5$ for right- and left-chiral particles.
[For the corresponding antiparticles the chemical potentials should be taken with the opposite sign, ${f}_{\bar{L}, \bar{R}}(\mathbf{k})=n_{F}(\epsilon_{k}\mp\mu_{5})$].
The left-right chirality imbalance is then characterized by the density of axial charge
\begin{equation}
 q_{5} = \int \frac{d^3\mathbf{k}}{(2\pi)^3} (f_{R} - f_{\bar{R}}- f_{L} + f_{\bar{L}})= 
 \frac{T^2 \mu_{5} }{3}
 \label{rhomu5}
\end{equation}
where in the last equality we assumed that $\mu_5 \ll T$.

The electron mass $m_e$ breaks the axial symmetry and thus the axial charge relaxes to zero: $\dot q_5 = - \Gamma_\flip q_5$ \cite{Note2}.
Assuming that the chirality relaxation is the slowest equilibration 
process in the plasma (we give a posterior justification of the assumption of the slowness of the chirality relaxation) the thermodynamic state~\eqref{fermi-dirac} with slowly varying $\mu_5 \neq 0$ can still be defined. We can then use Boltzmann's kinetic theory to compute $\Gamma_\flip$ as an asymptotic series in $m_e/T \ll 1$ \cite{Note3}. 

We now proceed to the calculation of the
chirality relaxation rate due to the $2\leftrightarrow 2$ processes of
Fig.~\ref{fig-Compton} within the framework of Boltzmann's kinetic theory. 
The rate of change of the axial charge due to the $2\leftrightarrow 2$ scattering processes is given by 
\begin{equation}
\label{rhodot}
 \dot q_5 = - \int \frac{d^3\mathbf{k}}{(2\pi)^3} (\mathcal{C}_{R} - \mathcal{C}_{\bar{R}}- 
 \mathcal{C}_{L} 
 + \mathcal{C}_{\bar{L}})
\end{equation}
where 
\begin{widetext}
\begin{multline}
\label{collision-term-general}
\mathcal{C}_{a}(\mathbf k) =\sum_{\{bcd\}}
\int\frac{d^{3}\mathbf{k}'}{(2\pi)^{3}}
\frac{d^{3}\mathbf{p}}{(2\pi)^{3}}
\frac{d^{3}\mathbf{p}'}{(2\pi)^{3}}
\frac{\left|\mathcal{M}^{ab}_{cd}(kp\to k'p')\right|^{2}}
{16\epsilon_{k}\epsilon_{k'}\epsilon_{p}\epsilon_{p'}}(2\pi)^{4}\delta^{(4)} (k+p-k'-p')
\\
\times\left[f_{a}(\mathbf{k})f_{b}(\mathbf{p})
(1\pm f_{c}(\mathbf{k}'))(1\pm f_{d}(\mathbf{p}'))-
(1\pm f_{a}(\mathbf{k}))(1\pm f_{b}(\mathbf{p}))f_{c}(\mathbf{k}')f_{d}(\mathbf{p}')\right],
\end{multline}
\end{widetext}
is Boltzmann's collision integral. In Eq.~\eqref{collision-term-general}, 
$k=(k^{0},\,\mathbf{k})$ is the 4-momentum, with $k^{0}=\epsilon_{k}=|\mathbf{k}|$ (the hard particles with $k\gtrsim T$ can be treated effectively as massless). The delta function takes into account the energy-momentum conservation in scattering. The subscripts $a,\,b,\,c,\,d$
run through the set of particle species $R,\,L,\, \bar R,\, \bar L,\, \gamma$; $f_a(\mathbf k)$ is the distribution function for 
the particle of type $a$ and in the expression $\mathbf \pm f_a(\mathbf k)$ the sign depends on the statistics of the particle $a$ (plus for a boson and minus for a fermion).
The amplitudes $\mathcal M^{ab}_{cd}$ are found by applying Feynman's rules to the diagrams shown in Fig.~\ref{fig-Compton}.

Expanding the thermal Fermi-Dirac distribution functions in the collision
integral on the right-hand side of Eq.~\eqref{rhodot} to the linear order in
$\mu_{5}$ and using Eq.~\eqref{rhomu5} we find that the chirality
imbalance decays exponentially
with the relaxation rate given by
\begin{widetext}
\begin{multline}
\label{Gamma-gen}
 \Gamma_{\rm flip} = \frac{3\pi}{T^3} 
 \int \frac{d^3\mathbf k}{(2\pi)^3} 
 \frac{d^3\mathbf p}{(2\pi)^3}\frac{d^3\mathbf q}{(2\pi)^3} 
 \Big\{n_{F}(k)n_F(p) [1+n_{B}(k')][1+n_{B}(p')] |\mathcal{M}_{\rm annih}|^{2} \\
 + n_{F}(k)n_B(p) [1+n_{B}(k')][1-n_{F}(p')] |\mathcal{M}_{\rm Compt}|^{2}\Big\} \frac{\delta(\epsilon_{k}+\epsilon_{p}-\epsilon_{k'}-\epsilon_{p'})}{\epsilon_{k}\epsilon_{k'}\epsilon_{p}\epsilon_{p'}},
\end{multline}
\end{widetext}
where $\mathbf k'=\mathbf k - \mathbf q$, $\mathbf p' = \mathbf p + \mathbf q$, and $n_{B}(p)=1/[\exp(\epsilon_{p}/T)-1]$ is the Bose-Einstein distribution function. We note, that in the weakly nonequilibrium situation $\mu_5\ll T$ considered here it is appropriate to take 
the Fermi-Dirac distribution functions and all matrix elements in Eq.~\eqref{Gamma-gen} at $\mu_5=0.$

Since we treat mass $m_{e}$ as a perturbation, we expand the matrix element of the Compton process as a perturbative series in $m_e$ and keep only the leading term
\begin{equation}
|\mathcal{M}^{(1)}|^{2}=\frac{8m_{e}^{2}e^{4} \epsilon_{k}\epsilon_{p}(1-\cos\theta_{kp})}{(q^{2})^{2}},
\end{equation}
where $\theta_{kp}$ is an angle between vectors $\mathbf{k}$ and $\mathbf{p}$. This matrix element contains a nonintegrable singularity at
$q=0$ which needs to be regularized by the environmental effects.  
To that end, we perform a partial resummation of the perturbative 
expansion in $\alpha$ to take into account the thermal self-energy corrections to the dispersion relations of quasiparticles:
\begin{equation}
\label{matrix-elem-thermal}
|\mathcal{M}^{(1), {\rm therm}}|^{2}=\frac{8m_{e}^{2}e^{4} \epsilon_{k}\epsilon_{p}(1-\cos\theta_{kp})}{|(q-\varpi)^{2}|^{2}},
\end{equation}
where $\varpi$ is a four-vector associated with the retarded self-energy of the intermediate particle by $\varpi^{\mu}= {\rm tr} (\gamma^\mu \Sigma_{\rm ret})/4$; see Refs.~\cite{Blaizot:1999xk,Blaizot:2001nr} and the discussion around Eq.~(A.3) in the Supplemental Material~\cite{Suppl} for more details. The presence of $\varpi$ regularizes the infrared divergence at $q\sim \sqrt{\alpha} T$.

Using the explicit expressions for the electron self-energy in the
HTL approximation,
we find that the chirality flipping rate
\begin{equation}
\label{gamma-flip-kinetic-final}
\Gamma_{\rm flip}=C\times \alpha \frac{m_{e}^{2}}{T},
\end{equation}
where the constant $C\approx 0.24$ (see Ref.~\cite{Suppl}, Sec.~A).

\begin{figure}[ht]
	\centering
	\includegraphics[width=0.7\linewidth]{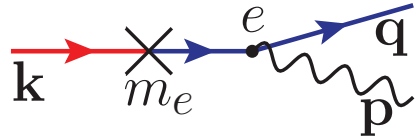}
	\caption{One of the $1\leftrightarrow 2$ collinear processes with chirality flip of the incoming electron (states with different chiralities are shown in different colors). Although for massless particles the process has a finite phase space, it is very sensitive to any modification of the particles' dispersion relations. This leads to an uncertainty in the probability of such a process which is addressed in Ref.~\cite{PaperII}.}
        \label{fig-lowest}
\end{figure}

Next, we briefly discuss other processes that contribute to the chirality flipping rate in the same order of perturbation theory as the Compton process. One such process is shown in Fig.~\ref{fig-lowest}. Its contribution 
to the chirality flipping rate can be estimated in a way similar to the $2\leftrightarrow 2$ case [see Eq.~(\ref{Gamma-gen})]
\begin{multline}
\label{Gamma-1to2}
    \Gamma_{\rm flip}^{1\leftrightarrow 2}\propto \frac{1}{T^{3}}\!\!\int\!\! d^3 \mathbf{k}\, d^3\mathbf{p}\, d^3\mathbf{q}\, n_{F}(\epsilon_{k})[1+n_{B}(\epsilon_{p})][1-n_{F}(\epsilon_{q})] \\ \times  \frac{\left|\mathcal{M}_{k\to pq}\right|^{2}}{\epsilon_{k}\epsilon_{q}\epsilon_{p}} \delta^{(3)}(\mathbf{k} - \mathbf{q} - \mathbf{p})\delta(\epsilon_k- \epsilon_q-\epsilon_p),
\end{multline}
where the matrix element reads as
\begin{equation}
\label{matrix-elem-1to2}
    \left|\mathcal{M}_{k\to pq}\right|^{2}=2e^{2}m_{e}^{2}\frac{k\cdot p}{k^{2}}.
\end{equation}
In vacuum, $\epsilon_k=|\mathbf{k}|$ and the process is only allowed for strictly collinear momenta of participating particles. Because of this kinematical constraint the process has an 
extremely unstable phase volume that can even be wiped out by an
infinitesimal deformation of the dispersion curves of the
particles. At the same time, 
the singularity of the matrix element \eqref{matrix-elem-1to2}
at $k=0$ leads to a nonintegrable divergence inside the available 
phase volume resulting in an uncertainty of ${0}/{0}$ type. The resolution of this uncertainty requires consideration of the finite lifetime of the particles involved in scattering as well as possible effects resulting from the multiple emission of soft photons~\cite{Baier:2000mf,Kovner:2003zj,Aurenche:2000gf,Arnold:2001ba,Arnold:2002ja,Arnold:2002zm}.
Such an analysis lies outside the scope of the present work. In Sec.~B of the Supplemental Material~\cite{Suppl}  
we explain why one should expect this contribution 
to the chirality flipping rate to be of the same parametric order as Eq.~(\ref{gamma-flip-kinetic-final}).

For details, we refer the interested reader to our companion paper, Ref.~\cite{PaperII}, 
where we investigate
the chirality flipping rate
within the framework of linear response theory.
The leading-order result for the chirality flipping rate derived in Ref.~\cite{PaperII} 
has the form given in Eq.~\eqref{gamma-flip-kinetic-final} with the coefficient $C$, which is a logarithmically varying function of $\alpha$. For $\alpha=1/137$ we find 
\begin{equation}
C \approx 1.17
\end{equation}
Thus, we find that the actual chirality flipping rate~\eqref{gamma-flip-kinetic-final} is 3 orders of magnitude as high as the previously used naive estimate $\Gamma_\flip^{\rm naive}$ (see, e.g., Ref.~\cite{Boyarsky:2011uy}).

\textit{Chirality flip across cosmic times.}---Our result~\eqref{gamma-flip-kinetic-final} enables us to compute the electron-mass induced the chirality flipping rate in the early Universe at temperatures $T \gtrsim m_e/\sqrt{\alpha}$; however, at much higher temperatures 
one should take into account other mechanisms responsible for 
chirality flipping.

At temperature above the electroweak phase transition the chirality flipping rate behaves as $\Gamma_{\rm flip}=(T_{R}/M_{\ast}) T$, where $M_{\ast}=M_{Pl}/(1.66\sqrt{g_{\ast}})$ and $T_{R}\sim 80\,{\rm TeV}$~\cite{Campbell:1992jd,Bodeker:2019ajh}. The responsible processes are various $2\leftrightarrow 2$ scatterings as well as the Higgs decay.
At temperatures well below the electroweak crossover, weak scatterings preserve chirality in the limit of zero masses of all fermions.
They are accompanied, however, by the subleading processes where chirality flips for one of the incoming or outgoing electrons with the probability proportional to $m_{e}^{2}/\langle p^{2}\rangle$.
The corresponding estimate for the reaction rate is given by
\begin{equation}
\label{EW-quadratic}
\Gamma_{\flip,\mathrm{\scriptscriptstyle EW}}\simeq G_{F}^{2}T^{5}\left(\frac{m_{e}}{3T}\right)^{2}.
\end{equation}
Unlike the QED case, there is no zero mass singularities because of the massive intermediate vector bosons.
There is also the contribution to the chirality flipping rate due to the Higgs (inverse) decay ($h \leftrightarrow e_L^- e_R^+$)
\begin{equation}
\label{rate-H}
\Gamma_{{\rm flip},H}=\frac{3\sqrt{2}}{\pi^{5}} G_{F}Tm_{e}^{2}\left(\frac{\pi m_{H}}{2T}\right)^{5/2}e^{-m_{H}/T},
\end{equation}
where $m_{H}$ is the Higgs boson mass.

\begin{figure}[t!]
	\centering
	\includegraphics[width=\linewidth]{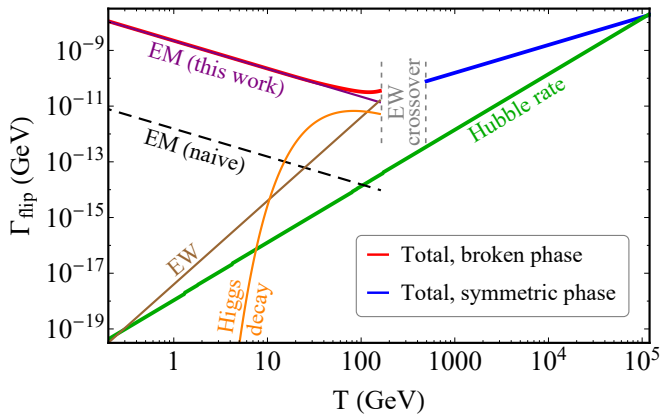}
	\caption{Chirality flipping rates due to different processes in comparison to the Hubble expansion rate $H(T)=T^2/M_{\ast}$ as functions of temperature. \label{fig-compare}}
\end{figure}

These results are summarized in Fig.~\ref{fig-compare}, which demonstrates that at temperatures $T \lesssim 80$~TeV $\Gamma_\flip$ always exceeds the Hubble expansion rate;  that the slowest $\Gamma_\flip(T)$ occurs at $T \simeq 100$~GeV and that below 100~GeV the ratio $\Gamma_\flip(T)/H(T) \gg 1$.

\textit{Conclusion and outlook.}---We have shown that the chirality flipping processes for electrons in QED plasma with $T \gg m_e$ occur much faster than one would naively expect:
it is proportional to the fine structure constant  $\alpha$, rather than $\propto\alpha^2$ (the latter dependence holds, for example, for chirality-preserving scatterings).
We used Boltzmann's collision integral to evaluate the contribution of the leading-order $2\leftrightarrow 2$ scattering processes (Fig.~\ref{fig-Compton}).
As $m_e/T \to 0$, the matrix elements for these processes exhibit the infrared singularity.
In order to obtain a meaningful result one has to proceed beyond tree-level analysis and invoke a partial resummation of the perturbation theory series.
In plasma such a resummation results in the singularity being regularized not by the mass $m_e$ but by the thermal mass of the electron $m_{\rm th} = \frac{eT}2$.
Our result in particular means that the chirality flipping rate is $\mathcal{O}(10^3)$ higher than was previously believed.

Chiral anomaly provides a coupling between the magnetic field and the axial 
current of electrons via the chiral magnetic effect~\cite{Vilenkin:1980fu,Joyce:1997uy,Alekseev:1998ds}.
Such a coupling has, in particular, been shown to lead to a special form of ``inverse cascade'' (transfer of the magnetic energy from smaller to larger scales) even in the absence of turbulence~\cite{Joyce:1997uy,Boyarsky:2011uy,Tashiro:2012mf,Hirono:2015rla,Rogachevskii:2017uyc,Brandenburg:2017rcb,Schober:2017cdw}. The inverse cascade is a remarkable example of macroscopic manifestation of a microscopic quantum effect.
This mechanism was, in particular, shown to increase the resilience of 
macroscopic magnetic fields against dissipative processes \cite{Boyarsky:2011uy, Brandenburg:2017rcb}. Chirality flipping suppresses the chiral magnetic effect, therefore it may switch off the inverse cascade before it completes the redistribution of energy between the electromagnetic modes. The present study shows that the accurate description of timescales associated with such counteracting mechanisms in a plasma requires a good microscopic understanding of the underlying quantum processes. Chirality flipping is not the only such mechanism. 
Recent  microscopic simulations~\cite{Figueroa:2017qmv,Figueroa:2017hun,Figueroa:2019jsi} hint that the anomaly induced rate of redistribution of energy between the electromagnetic modes may significantly exceed its classical estimate, presumably due to quantum effects arising at short length scales. These findings call for 
further revision of the MHD of axially charged plasmas based 
on a first-principles approach along the lines of the present study.

We are grateful to Artem Ivashko, Oleksandr Gamayun, Kyrylo Bondarenko, Alexander Monin, and Mikhail Shaposhnikov for valuable discussions.
This project has received funding from the European Research Council (ERC) under the European Union's Horizon 2020 research and innovation programme (GA 694896) and from the Carlsberg Foundation. The work of O.~S. was supported by the Swiss National Science Foundation Grant No. 200020B\_182864.

\clearpage
\appendix

\onecolumngrid
\section*{Supplemental Material}

\renewcommand{\theequation}{\thesection.\arabic{equation}}
\renewcommand{\thefigure}{\thesection.\arabic{figure}}
\numberwithin{equation}{section}

\section{Collision integral involving $2\leftrightarrow 2$ processes}
\label{app-collision}

In this section we calculate the collision integrals corresponding to the processes of the Compton scattering and electron-positron annihilation with the flip of chirality. Let us consider these processes separately.

\paragraph{The Compton scattering.} The matrix element of the $t$-channel process $e_{L}(k)+\gamma(p)\to \gamma(k')+e_{R}(p')$ shown in Fig.~1(a) reads as
\begin{equation}
\mathcal{M}_{C}=\frac{1}{i}(-ie)^{2} \bar{u}_{s'}(p')P_{R}\gamma^{\nu} i\mathcal{S}_{\rm ret}(q)\gamma^{\mu}P_{R}u_{s}(k)\varepsilon^{*}_{\mu}(k',\lambda')\varepsilon_{\nu}(p,\lambda),
\end{equation}
where $P_{R}=(1+\gamma_{5})/2$ is the right chiral projector and the propagator has the form
\begin{equation}
\label{prop-fermion}
\mathcal{S}_{\rm ret}(q)=[\mathcal{S}_{0,{\rm ret}}(q)-\Sigma_{\rm ret}(q)]^{-1}=\frac{\cancel{q}-\cancel{\varpi}+m_{e}\mathds{1}}{(q-\varpi)^{2}-m_{e}^{2}}.
\end{equation}
The 4-vector $\varpi^{\mu}={\rm tr}(\gamma^{\mu}\Sigma_{\rm ret})/4$ equals \cite{LeBellac}
\begin{equation}
\label{Sigma-mu}
\varpi^{\mu}=\left(\varpi^{0},\,\frac{\mathbf{q}}{q} \left[-\frac{m_{\rm th}^{2}}{2q}+\frac{q^{0}}{q}\varpi^{0}\right]\right),
\end{equation}
where $m_{\rm th}=eT/2$ is the electron thermal mass and the function $\varpi^{0}$ is given by
\begin{equation}
\label{sigma-0-complex}
\varpi^{0}(q^{0},\mathbf{q})=\frac{m_{\rm th}^{2}}{2}\int\frac{d\Omega_{\mathbf{v}}}{4\pi}\frac{1}{q^{0}-\mathbf{q}\cdot\mathbf{v}}=\frac{m_{\rm th}^{2}}{4k}\ln\frac{q^{0}+q}{q^{0}-q}.
\end{equation}
Note, that we use the retarded propagator for the intermediate particle while computing the scattering matrix element for the collision integral. This prescription arises in finite temperature quantum field theory when one derives the collision term from the Schwinger-Keldysh formalism, see \cite{Blaizot:1999xk,Blaizot:2001nr} as well as \cite{Ghiglieri:2020dpq} for a recent discussion.

Because of the chiral projectors, only the massive term in the numerator in Eq.~(\ref{prop-fermion}) contributes to the matrix element. In the leading order in $m_{e}$ we get
\begin{equation}
\mathcal{M}_{C}^{(1)}=-m_{e}e^{2}\frac{\bar{u}_{s'}(p')\gamma^{\nu}\gamma^{\mu}P_{R}u_{s}(k)}{(q-\varpi)^{2}}\varepsilon^{*}_{\mu}(k',\lambda')\varepsilon_{\nu}(p,\lambda).
\end{equation}
Taking the squared modulus of this matrix element and summing over all possible spins and polarizations (we sum over the spin projections because we included the chiral projectors directly in the matrix element), we get
\begin{equation}
\label{matrix-elem-squre-compton}
|\mathcal{M}_{C}^{(1)}|^{2}=\frac{8m_{e}^{2}e^{4}(k\cdot p')}{|(q-\varpi)^{2}|^{2}}.
\end{equation}

\paragraph{The annihilation process.} Let us now consider the process of annihilation $e_{L}(k)+\overline{e_{R}}(p)\to \gamma(k')+\gamma(p')$ shown in Fig.~1(b). It is worth noting that the incoming positron is the antiparticle to the right electron, i.e., it is left, so that the chirality is not conserved in this reaction. Following the same steps as in the case of Compton scattering, we obtain the matrix element
\begin{equation}
\mathcal{M}_{A}^{(1)}=-m_{e}e^{2}\frac{\bar{v}_{s'}(p)\gamma^{\nu}\gamma^{\mu}P_{R}u_{s}(k)}{(q-\varpi)^{2}}\varepsilon^{*}_{\mu}(k',\lambda')\varepsilon^{*}_{\nu}(p',\lambda)
\end{equation}
and its squared modulus
\begin{equation}
\label{matrix-elem-squre-annihil}
|\mathcal{M}_{A}^{(1)}|^{2}=\frac{8m_{e}^{2}e^{4}(k\cdot p)}{|(q-\varpi)^{2}|^{2}}, \qquad q=k-k'.
\end{equation}
This matrix element coincides with that of the Compton process (\ref{matrix-elem-squre-compton}) up to the terms $\mathcal{O}(q)$ in the numerator. It is important to note that there is also the $u$-channel of annihilation when the outcoming photons are interchanged. However, the matrix element is exactly the same with $q=k-p'$ instead of $q=k-k'$. Changing the variables $k'\leftrightarrow p'$ in the collision integral we can see that the result is simply twice the result of $t$-channel. There is, however, the factor $1/2$ in front of the collision integral which takes into account the indistinguishability of the outcoming photons. In our calculation, we omit both, the $u$-channel and the factor of $1/2$. 

Taking into account the identity
\begin{equation}
[1-n_{F}(p)]n_{B}(p)=n_{F}(p)[1+n_{B}(p)]=\frac{1}{2\,{\rm sinh\,}p/T},
\end{equation}
we conclude that the Compton scattering and the annihilation process make equal contributions to the collision integral.

\paragraph{Calculation of the chirality flipping rate.} Substituting the expressions (\ref{matrix-elem-squre-compton}) and (\ref{matrix-elem-squre-annihil}) for to the Compton scattering and annihilation processes into the expression for chirality flipping rate~(6), we arrive at the following expression 
\begin{equation}
\label{chirality-flip-compton-1}
\Gamma_{\rm flip}^{2\leftrightarrow 2}=\frac{3m_{e}^{2}e^{4}T}{128\pi} \int_{0}^{\infty}q\,dq \int_{0}^{\pi}d\cos\theta_{kq}\left.\frac{ 1-\cos^{2}\theta_{kq}}{|(q^{0}-\varpi^{0})^{2}-(\mathbf{q}-\boldsymbol{\varpi})^{2}|^{2}}\right|_{q^{0}=q\cos\theta_{kq}}.
\end{equation}

Let us carefully consider its denominator
\begin{eqnarray}
\psi(q,\cos\theta_{kq})&=&\left.\frac{(q^{0}-\varpi^{0})^{2}-(\mathbf{q}-\boldsymbol{\varpi})^{2}}{q^{2}}\right|_{q^{0}=q\cos\theta_{kq}}\nonumber\\
&=&\Big[\cos\theta_{kq}-\frac{m_{\rm th}^{2}}{4q^{2}}\Big(\ln\frac{1+\cos\theta_{kq}}{1-\cos\theta_{kq}}-i\pi\Big)\Big]^{2}\nonumber\\
&-&\Big[1+\frac{m_{\rm th}^{2}}{2q^{2}}-\cos\theta_{kq}\frac{m_{\rm th}^{2}}{4q^{2}}\Big(\ln\frac{1+\cos\theta_{kq}}{1-\cos\theta_{kq}}-i\pi\Big)\Big]^{2}
.
\end{eqnarray}
It is easy to see that it depends only on $q^{2}/m_{\rm th}^{2}$ and satisfies
\begin{equation}
\psi(q,-\cos\theta_{kq})=\psi^{*}(q,\cos\theta_{kq}),
\end{equation}
so that $|\psi|^{2}$ is invariant under the reflection $\theta_{kq}\to \pi -\theta_{kq}$. 
Introducing the new integration variables
\begin{equation}
\xi=q^{2}/m_{\rm th}^{2},\qquad y=\cos\theta_{kq},
\end{equation}
we get the expression for the chirality flipping rate in the form
\begin{equation}
\Gamma_{\rm flip}^{2\leftrightarrow 2}=\frac{m_{e}^{2}}{T}\alpha\times\frac{3}{8}\int_{0}^{\infty}d\xi\int_{0}^{1}dy\frac{1-y^{2}}{\xi^{2}|\psi(m_{\rm th}\sqrt{\xi},y)|^{2}}.
\end{equation}
As the final step, we show that
\begin{eqnarray}
&&\frac{1-y^{2}}{\xi^{2}|\psi(m_{\rm th}\sqrt{\xi},y)|^{2}}\nonumber\\
&&=\frac{1-y^{2}}{\xi^{2}\left|\Big[y-\frac{1}{4\xi}\Big(\ln\frac{1+y}{1-y}-i\pi\Big)\Big]^{2}-\Big[1+\frac{1}{2\xi}-\frac{y}{4\xi}\Big(\ln\frac{1+y}{1-y}-i\pi\Big)\Big]^{2}\right|^{2}}\nonumber\\
&&=\frac{\xi^{2}/(1-y^{2})}{\left[\left(\xi+\frac{1}{4}\ln\frac{1+y}{1-y}+\frac{1}{2(1-y)}\right)^{2}+\frac{\pi^{2}}{16}\right]\left[\left(\xi-\frac{1}{4}\ln\frac{1+y}{1-y}+\frac{1}{2(1+y)}\right)^{2}+\frac{\pi^{2}}{16}\right]},
\end{eqnarray}
and we end up with Eq.~(9) where the constant $C$ equals to
\begin{eqnarray}
C&=&\frac{3}{8}\int_{0}^{1}\frac{dy}{1-y^{2}}\int_{0}^{\infty} \frac{\xi^{2}\,d\xi}{\left[\left(\xi+\frac{1}{4}\ln\frac{1+y}{1-y}+\frac{1}{2(1-y)}\right)^{2}+\frac{\pi^{2}}{16}\right]\left[\left(\xi-\frac{1}{4}\ln\frac{1+y}{1-y}+\frac{1}{2(1+y)}\right)^{2}+\frac{\pi^{2}}{16}\right]}\nonumber\\
&\approx& 0.24.
\end{eqnarray}

\section{Chirality flipping rate from $1\leftrightarrow 2$ processes}
\label{app-1to2}

The contribution to the chirality flipping rate from the $1\leftrightarrow 2$ process shown in Fig.~2 can be estimated by Eq.~(10). Let us take into account the thermal corrections and show that they lead to the finite answer. For further convenience, let us decompose the momenta into components along the momentum $\mathbf{k}$ of the incoming electron and transverse to it,
\begin{equation}
    \mathbf{p}=p_{\parallel}\hat{\mathbf{k}}+\mathbf{p}_{\perp},\ \ \mathbf{q}=(k-p_{\parallel})\hat{\mathbf{k}}-\mathbf{p}_{\perp},
\end{equation}
where $\hat{\mathbf{k}}=\mathbf{k}/k$ and in the second expression we used the momentum conservation law.
Treating the longitudinal components of all momenta to be $\sim T$ (we will see that this is true \textit{a posteriori}), we can expand the dispersion relations as follows
\begin{equation}
\epsilon_{k}\approx k+\frac{m_{\rm th}^{2}}{2k},\quad
\epsilon_{q}\approx k-p_{\parallel}+\frac{m_{\rm th}^{2}+p_{\perp}^{2}}{2(k-p_{\parallel})},\quad 
\epsilon_{p}\approx p_{\parallel}+\frac{m_{\gamma}^{2}+p_{\perp}^{2}}{2p_{\parallel}}.
\end{equation}
Here $m_{\rm th}=eT/2$ and $m_{\gamma}=eT/\sqrt{6}$ are the asymptotic thermal masses of the electron and photon, respectively \cite{LeBellac}.

The HTL effective theory predicts the modification of the dispersion relations. 
However, this would immediately wipe out all the available phase space for $1\leftrightarrow 2$ processes and lead to the vanishing contribution to the chirality flipping rate. Fortunately, the higher order (beyond HTL) corrections give rise also to the finite decay width of the quasiparticles \cite{Thoma:1995ju,Blaizot:1996az,Blaizot:1996hd} and allow for a slight violation of the energy conservation in the collision event. The electron decay width equals to $\gamma_{e}\approx e^{2}T/(4\pi) \log e^{-1}$, while the photon decay width is of higher order in $e$ and thus can be neglected. At technical level, we can incorporate this finite decay width by replacing the delta function of energies in Eq.~(10) by the corresponding Lorentz contour of the width $2\gamma_{e}$.

The Lorentz function works only when its argument is less or of the order its width, 
\begin{equation}
|\epsilon_{k}-\epsilon_{q}-\epsilon_{p}|\approx \frac{m_{\gamma}^{2}}{2p_{\parallel}}+\frac{m_{\rm th}^{2} p_{\parallel}}{2k(k-p_{\parallel})}+\frac{p_{\perp}^{2} k}{2p_{\parallel}(k-p_{\parallel})}\lesssim 2\gamma_{e}\sim Te^{2}\log e^{-1}.
\end{equation}
This immediately gives the restrictions on the longitudinal and transverse components of the momenta
\begin{equation}
\label{constraints-momenta}
k>p_{\parallel}\gtrsim m_{\rm th}^{2}/\gamma_{e}\sim T/\log e^{-1}, \qquad p_{\perp} \lesssim m_{\rm th}.
\end{equation}

Now, let us consider the matrix element (11). The scalar products equal to
\begin{equation}
k\cdot p\approx \frac{k(m_{\gamma}^{2}+p_{\perp}^{2})}{2p_{\parallel}}+\frac{p_{\parallel}m_{\rm th}^{2}}{2k},\quad k^{2}\approx m_{\rm th}^{2}.
\end{equation}
Taking into account constraints (\ref{constraints-momenta}), we obtain that 
\begin{equation}
\left|\mathcal{M}_{k\to pq}\right|^{2}=\mathcal{O}(1)\times e^{2}m_{e}^{2}.
\end{equation}
Then, the chirality flipping rate can be written as follows
\begin{multline}
    \Gamma_{\rm flip}^{1\leftrightarrow 2}\propto \frac{e^{2}m_{e}^{2}}{T^{3}}\int_{0}^{\infty}\!\!k^{2}\,dk \int_{0}^{k}\!dp_{\parallel} \frac{n_{F}(k)[1+n_{B}(p_{\parallel})][1-n_{F}(k-p_{\parallel})]}{k p_{\parallel}(k-p_{\parallel})}\\ \times \int p_{\perp}\,dp_{\perp} \delta_{2\gamma_{e}}\left(\frac{m_{\gamma}^{2}}{2p_{\parallel}}+\frac{m_{\rm th}^{2} p_{\parallel}}{2k(k-p_{\parallel})}+\frac{p_{\perp}^{2} k}{2p_{\parallel}(k-p_{\parallel})}\right)\\
    \propto \frac{e^{2}m_{e}^{2}}{T^{3}}\int_{0}^{\infty}\!\!dk \int_{0}^{k}\!dp_{\parallel}\  n_{F}(k)[1+n_{B}(p_{\parallel})][1-n_{F}(k-p_{\parallel})] \frac{2}{\pi}{\rm arctan}\frac{4\gamma_{e}}{\frac{m_{\gamma}^{2}}{p_{\parallel}}+\frac{m_{\rm th}^{2}p_{\parallel}}{k(k-p_{\parallel})}}.
\end{multline}
Without the arctangent, the integral over $p_{\parallel}$ would be logarithmically divergent for small momenta because of the Bose-Einstein distribution function. However, at the scale $p_{\parallel,\rm min}\sim m_{\gamma}^{2}/\gamma_{e}\sim T/\log e^{-1}$ the arctangent cuts this divergence. Finally, we get the estimate
\begin{equation}
\Gamma_{\rm flip}^{1\leftrightarrow 2}\sim \frac{m_{e}^{2}}{T}\times \alpha \log\log\alpha^{-1},
\end{equation}
which is again of the first order in the electromagnetic coupling constant with a slight logarithmic enhancement. Thus, we confirm that the nearly collinear $1\leftrightarrow 2$ processes also contribute to the leading order chirality flipping rate. This contribution is studied in our companion paper \cite{PaperII}.

\end{document}